\documentclass[aps,prb,showpacs,superscriptaddress,twocolumn]{revtex4}
\usepackage{amsmath}
\usepackage{amsfonts}
\usepackage{graphicx}
\usepackage{subfigure}
\usepackage{natbib}

\begin{document}

\title{Magnetoconductance properties of rectangular arrays of spintronic
quantum rings}
\author{Orsolya K\'{a}lm\'{a}n}
\affiliation{Department of Nonlinear and Quantum Optics, Research Institute for Solid
State Physics and Optics,Hungarian Academy of Sciences, Konkoly-Thege Mikl%
\'{o}s \'{u}t 29-33, H-1121 Budapest, Hungary}
\affiliation{Department of Theoretical Physics, University of Szeged, Tisza Lajos k\"{o}r%
\'{u}t 84, H-6720 Szeged, Hungary}
\author{P\'{e}ter F\"{o}ldi}
\affiliation{Department of Theoretical Physics, University of Szeged, Tisza Lajos k\"{o}r%
\'{u}t 84, H-6720 Szeged, Hungary}
\author{Mih\'{a}ly G. Benedict}
\email{benedict@physx.u-szeged.hu}
\affiliation{Department of Theoretical Physics, University of Szeged, Tisza Lajos k\"{o}r%
\'{u}t 84, H-6720 Szeged, Hungary}
\author{F. M. Peeters}
\affiliation{Departement Fysica, Universiteit Antwerpen, Groenenborgerlaan 171, B-2020
Antwerpen, Belgium}

\begin{abstract}
Electron transport through multi-terminal rectangular arrays of quantum
rings is studied in the presence of Rashba-type spin-orbit interaction (SOI)
and of a perpendicular magnetic field. Using the analytic expressions for
the transmission and reflection coefficients for single rings we obtain the
conductance through such arrays as a function of the SOI strength, the
magnetic flux, and of the wave vector $k$ of the incident electron. Due to
destructive or constructive spin interferences caused by the SOI, the array
can be totally opaque for certain ranges of $k$, while there are parameter
values where it is completely transparent. Spin resolved transmission
probabilities show nontrivial spin transformations at the outputs of the
arrays. When point-like random scattering centers are placed between the
rings, the Aharonov-Bohm peaks split, and an oscillatory behavior of the
conductance emerges as a function of the SOI strength.
\end{abstract}

\maketitle

\section{Introduction}

Magnetoconductance oscillations of quantum rings made of semiconducting
materials \cite{KTHS06} exhibiting Rashba-type spin-orbit interaction\cite%
{G00,NATE97,SKY01} (SOI) have been intensely studied in the past few years.
These effects are manifestations of flux- and spin-dependent quantum
interference phenomena. In view of the possible spintronic applications and
the conceptual importance of these interference effects in
multiply-connected domains, closed single quantum rings (without attached
leads),~\cite{SGZ03,SC06,Y06,YCSX07b} as well as two- or three-terminal ones were
investigated \cite%
{BIA84,NMT99,MPV04,FR04,FMBP05,WV05,SN05,KMGA05,BO07,SP05,VKPB07,YPS03,FHR01,CHR07}
extensively. Additionally, the conductance properties of a linear chain of
rings have also been determined.~\cite{MVP05}

In this paper we present a method that enables one to calculate the conductance
and the spin transport properties of two-dimensional rectangular
arrays consisting of quantum rings with Rashba-type SOI \cite{R60} and with
a perpendicular magnetic field. Such arrays, fabricated from e.g. an
InAlAs/InGaAs based 2DEG,~\cite{KNAT02} have been studied in a recent
experiment \cite{BKSN06} and in a subsequent theoretical work \cite{ZW07} to
demonstrate the time-reversal Aharonov-Casher effect.~\cite{AC84} Here we
present a more general survey of the magnetoconductance properties of such
devices, including the perturbative treatment of the magnetic field which
still allows us to analytically solve the scattering problem in case of
two-, three- and four-terminal rings, which are then used as building blocks
of larger arrays. Our method is based on analytic results, and can be used
for an arbitrary configuration. For the sake of definiteness, we consider 3$%
\times $3, 4$\times $4 and 5$\times $5 rectangular arrays,~\cite{BKSN06,ZW07}
which are closed in the vertical, and open in the horizontal direction.
Additionally, we study the magnetoconductance properties and spin resolved
transmission probabilities of the same array geometry with only one input
channel. We also investigate to what extent the conductance properties are
modified by the presence of point-like random scattering centers between the
rings. In our calculations we assume that the rings are narrow enough to be
considered one-dimensional and the transport of the electrons through the
arrays is ballistic. We determine the magnetoconductance in the framework of
the Landauer-B\"{u}ttiker formalism.~\cite{D95}

Rectangular arrays~\cite{BKSN06,ZW07} --depending on the number of input
leads-- consist of two-, three- and four-terminal rings (see Fig.~\ref{fig1}%
), where the two- and three-terminal ones are situated on the boundary of
the arrays as shown in Fig.~\ref{networkfig} with or without the input leads
displayed by dashed lines. The transmission and reflection properties of two- and
three-terminal rings have been determined in previous works \cite%
{NMT99,MPV04,FR04,FMBP05a,WV05,AL93,X92,FKBP06,KFB06,KFBP08} but the effect
of the magnetic field on the spin degree of freedom has not been taken into
account for an arbitrary geometry. Additionally, the most general boundary 
condition that is required
by this two-dimensional problem has not been investigated so far. Therefore
in Sec.~\ref{modelsec} we first consider a perpendicular magnetic field as a
weak perturbation, then, in order to account for all possible reflections
and transmissions when building up the array from single rings, we
generalize our previous results to the case when electrons can enter/exit on
any of the terminals of a three-terminal ring (results for two- and
four-terminal rings are presented in the Appendix). Next, in Sec.~\ref%
{arraysecsubsec1} the individual rings are used as building blocks of the
arrays by fitting the wave functions and their derivatives in the points
where neighboring rings touch each other. Magnetoconductance properties are
presented here as a function of the wave number $k$ of the incoming
electron, the magnetic flux and the SOI strength. Spin resolved transmission
probabilities on the output side of the arrays are also derived. In Sec.~\ref%
{arraysecsubsec2} we investigate the effect of random Dirac-delta scattering
potentials in between the rings.

\bigskip

\section{Building blocks of two-dimensional arrays: Single quantum rings}

\label{modelsec} In this section we consider a single, narrow quantum ring 
\cite{AL93} of radius $a$ located in the $xy$ plane in the presence of
Rashba SOI \cite{R60} and a perpendicular magnetic field $B$. If $B$ is
relatively weak, then the interaction between the electron spin and the
field, i.e. the Zeeman term can be treated as a perturbation and the
relevant dimensionless Hamiltonian reads \cite{MMK02,MPV04}%
\begin{equation}
H=\left[ \left( -i\frac{\partial }{\partial \varphi }-\frac{\Phi }{\Phi _{0}}%
+\frac{\omega _{\text{SO}}}{2\Omega }\sigma _{r}\right) ^{2}-\frac{\omega _{%
\text{SO}}^{2}}{4\Omega ^{2}}\right] +H_{\text{p}},  \label{Ham}
\end{equation}%
where $\varphi $ is the azimuthal angle of a point on the ring, $\Phi $
denotes the magnetic flux encircled by the ring, $\Phi _{0}=h/e$ is the unit
flux, and $\omega _{\text{SO}}=\alpha /\hbar a$ is the frequency associated
with the spin-orbit interaction. $\hbar \Omega =\hbar ^{2}/2m^{\ast }a^{2}$
characterizes the kinetic energy with \thinspace $m^{\ast }$ being the
effective mass of the electron, and the radial spin operator is given by $%
\sigma _{r}=\sigma _{x}\cos \varphi +\sigma _{y}\sin \varphi $. The
perturbative term $H_{\text{p}}$ is given by \cite{MPV04}%
\begin{equation*}
H_{\text{p}}=\frac{\omega _{\text{L}}}{\Omega }\sigma _{z},
\end{equation*}%
where $\omega _{\text{L}}=g^{\ast }eB/4m$ with $g^{\ast }$ and $m$ being the
effective gyromagnetic ratio and the free electron mass, respectively.

The energy eigenvalues of the unperturbed Hamiltonian are 
%\label{eigenvalvect}
\begin{subequations}
\begin{equation}
E_{0}^{\left( \mu \right) }\!(\kappa )=\!\left( \!\kappa -\frac{\Phi }{\Phi
_{0}}\!\right) ^{\!2}+\left( -1\right) ^{\mu }\!\left( \!\kappa -\frac{\Phi 
}{\Phi _{0}}\!\right) w+\frac{1}{4},\ \left( \mu =1,2\right) ,
\label{eigenval}
\end{equation}%
and the corresponding eigenvectors in the $|\uparrow_z \rangle
,|\downarrow_z \rangle $ eigenbasis of $\sigma _{z}$ read 
\begin{equation}
\psi ^{\left( \mu \right) }(\kappa ,\varphi )=e^{i\kappa \varphi }%
\begin{pmatrix}
e^{-i\varphi /2}u^{\left( \mu \right) } \\ 
e^{i\varphi /2}v^{\left( \mu \right) }%
\end{pmatrix}%
,  \label{u1v1}
\end{equation}%
where $u^{\left( 1\right) }\!=\!-v^{\left( 2\right) }\!=\!\cos (\theta /2)$; 
$u^{\left( 2\right) }\!=\!v^{\left( 1\right) }\!=\!\sin (\theta /2)$ and 
\end{subequations}
\begin{equation}
\tan (\theta /2)=\frac{\Omega }{\omega _{\text{SO}}}\left( 1-w\right) ,
\label{theta}
\end{equation}%
with $w=\sqrt{1+\omega _{\text{SO}}^{2}/\Omega ^{2}}$.

The matrix elements of $H_{\text{p}}$ in the basis of these eigenstates are
obtained as 
\begin{equation*}
\left\langle \psi ^{\left( \mu \right) }\right\vert H_{\text{p}}\left\vert
\psi ^{\left( \mu \right) }\right\rangle =\left( -1\right) ^{\mu +1}\frac{%
\omega _{\text{L}}}{\Omega }\cos \theta =\left( -1\right) ^{\mu +1}\frac{%
\omega _{\text{L}}}{\Omega }\frac{1}{w},
\end{equation*}%
\begin{equation*}
\left\langle \psi ^{\left( 1\right) }\right\vert H_{\text{p}}\left\vert \psi
^{\left( 2\right) }\right\rangle =\frac{\omega _{\text{L}}}{\Omega }\sin
\theta .
\end{equation*}%
In the first-order approximation one neglects the off-diagonal elements;
this is reasonable if they are small, i.e., if $\omega _{\text{L}}/\Omega
\ll k^{2}a^{2}$, where $k$ denotes the wave number of the incident electron,
which is described as a plane wave. Within this approximation, the
eigenspinors are not perturbed and their direction is still specified by the
angle $\theta $ given by Eq.~(\ref{theta}). The energy eigenvalues including
the first-order corrections are given by 
\begin{equation*}
E^{\left( \mu \right) }\!(\kappa )=E_{0}^{\left( \mu \right) }\!(\kappa
)+\left( -1\right) ^{\mu +1}\frac{\omega _{\text{L}}}{\Omega }\frac{1}{w}.
\end{equation*}

Imposing the condition of energy conservation $k^{2}\!a^{2}=E^{\left( \mu
\right) }\!(\kappa )$ determines the possible values of $\kappa $:%
\begin{equation*}
\kappa _{j}^{\left( \mu \right) }=\left( -1\right) ^{\mu +1}\left[ \frac{w}{2%
}+\left( -1\right) ^{j}q^{\left( \mu \right) }\right] +\frac{\Phi }{\Phi _{0}%
},
\end{equation*}%
where $\mu ,j=1,2$ and 
\begin{equation}
q^{\left( \mu \right) }=\sqrt{q^{2}+\left( -1\right) ^{\mu }\frac{\omega _{%
\text{L}}}{\Omega }\frac{1}{w}}  \label{q_mu}
\end{equation}
with $q=\sqrt{(\omega _{\text{SO}}/2\Omega )^{2}+E/\hbar \Omega }$, and $%
E=\hbar ^{2}k^{2}/2m^{\ast }$ denoting the energy of the incoming electron.
The corresponding four eigenspinors read 
\begin{equation}
\psi _{j}^{\left( 1\right) }(\kappa _{j}^{\left( 1\right) },\varphi
)=e^{i\kappa _{j}^{\left( 1\right) }\varphi }%
\begin{pmatrix}
e^{-i\varphi /2}\cos (\theta /2) \\ 
e^{i\varphi /2}\sin (\theta /2)%
\end{pmatrix}%
,
\end{equation}%
\begin{equation}
\psi _{j}^{\left( 2\right) }(\kappa _{j}^{\left( 2\right) },\varphi
)=e^{i\kappa _{j}^{\left( 2\right) }\varphi }%
\begin{pmatrix}
e^{-i\varphi /2}\sin (\theta /2) \\ 
-e^{i\varphi /2}\cos (\theta /2)%
\end{pmatrix}%
.
\end{equation}%
The wave functions belonging to the same energy in the different sections of
the ring are linear combinations of these eigenspinors.

%%%%%%%%%%%%%%%%%%%%%%%%%%%%%%%%%%%%%%%%%%%%%%%%%%%%%%%%%%%%%%%%%%%%%%%%%%%%%%%%%%%
\begin{figure}[tbh]
\centering
\subfigure{
\label{fig1a}
\includegraphics[width=4.0cm]{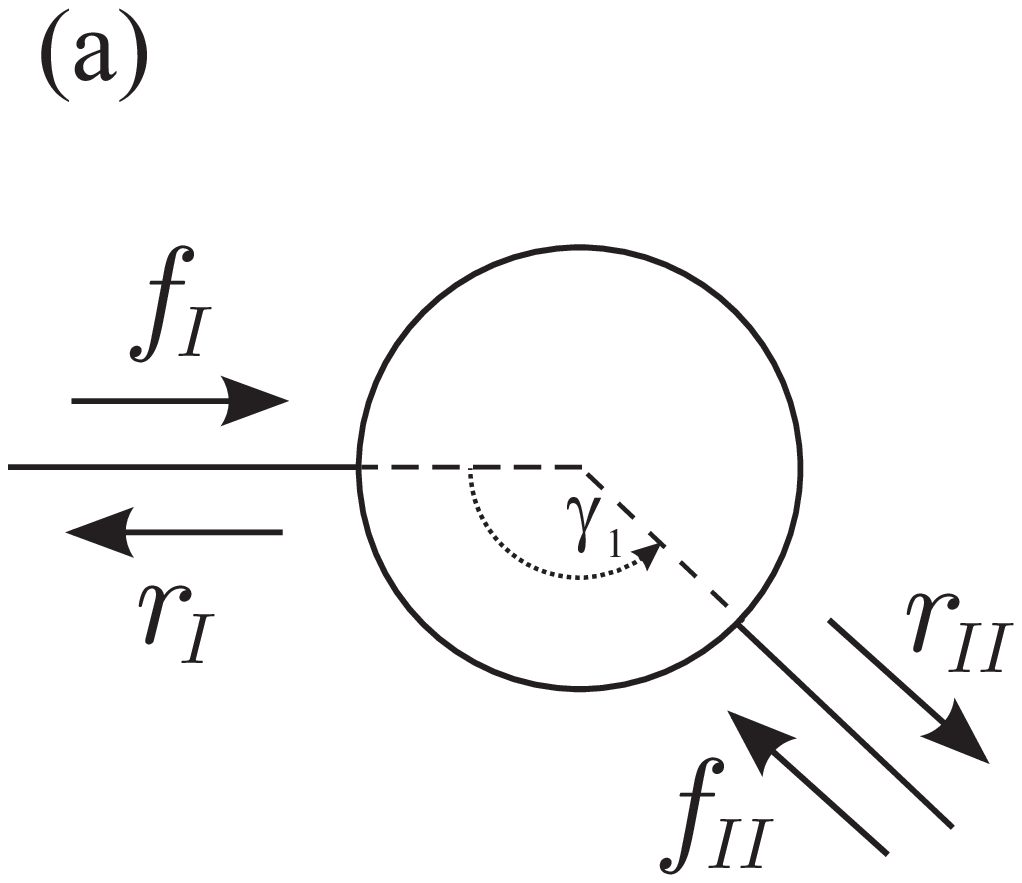}}\hspace{0.0cm} 
\subfigure{
\label{fig1b}
\includegraphics[width=4.0cm]{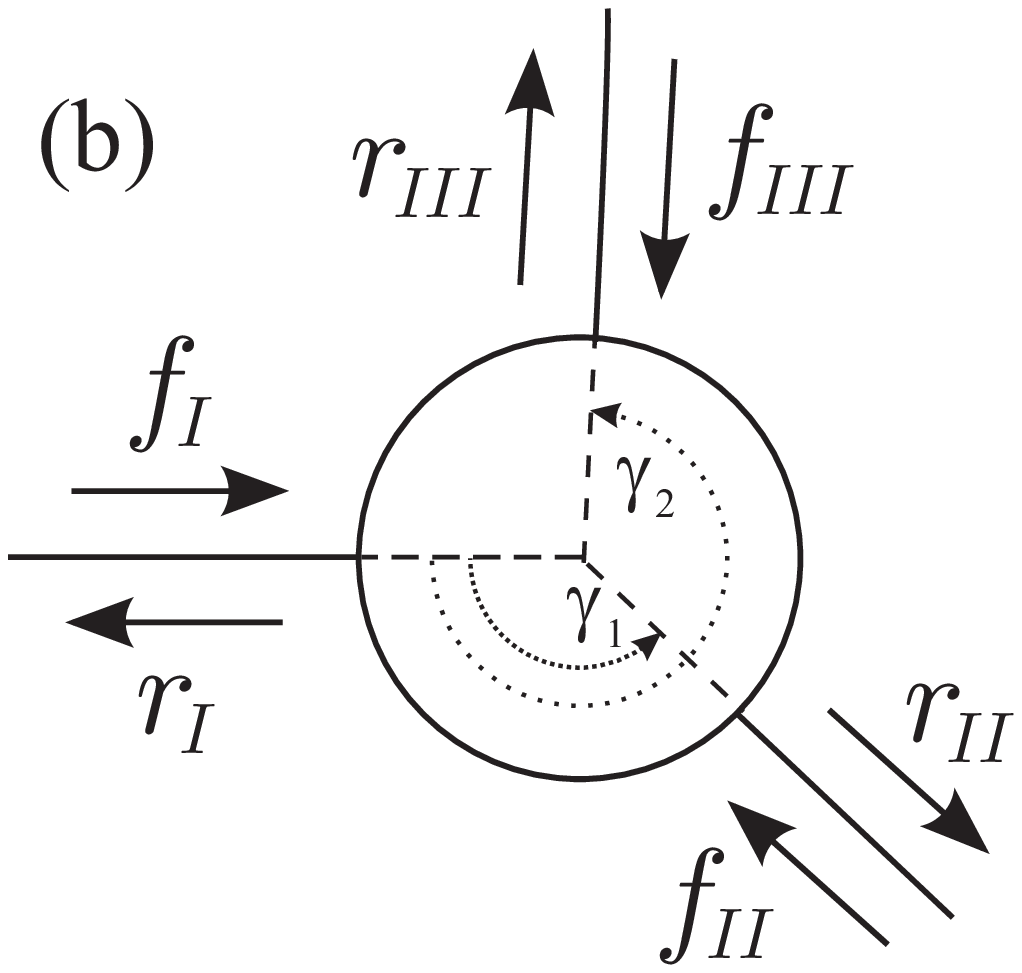}}\hspace{0.0cm}%
\newline
\subfigure{
\label{fig1c}
\includegraphics[width=5.0cm]{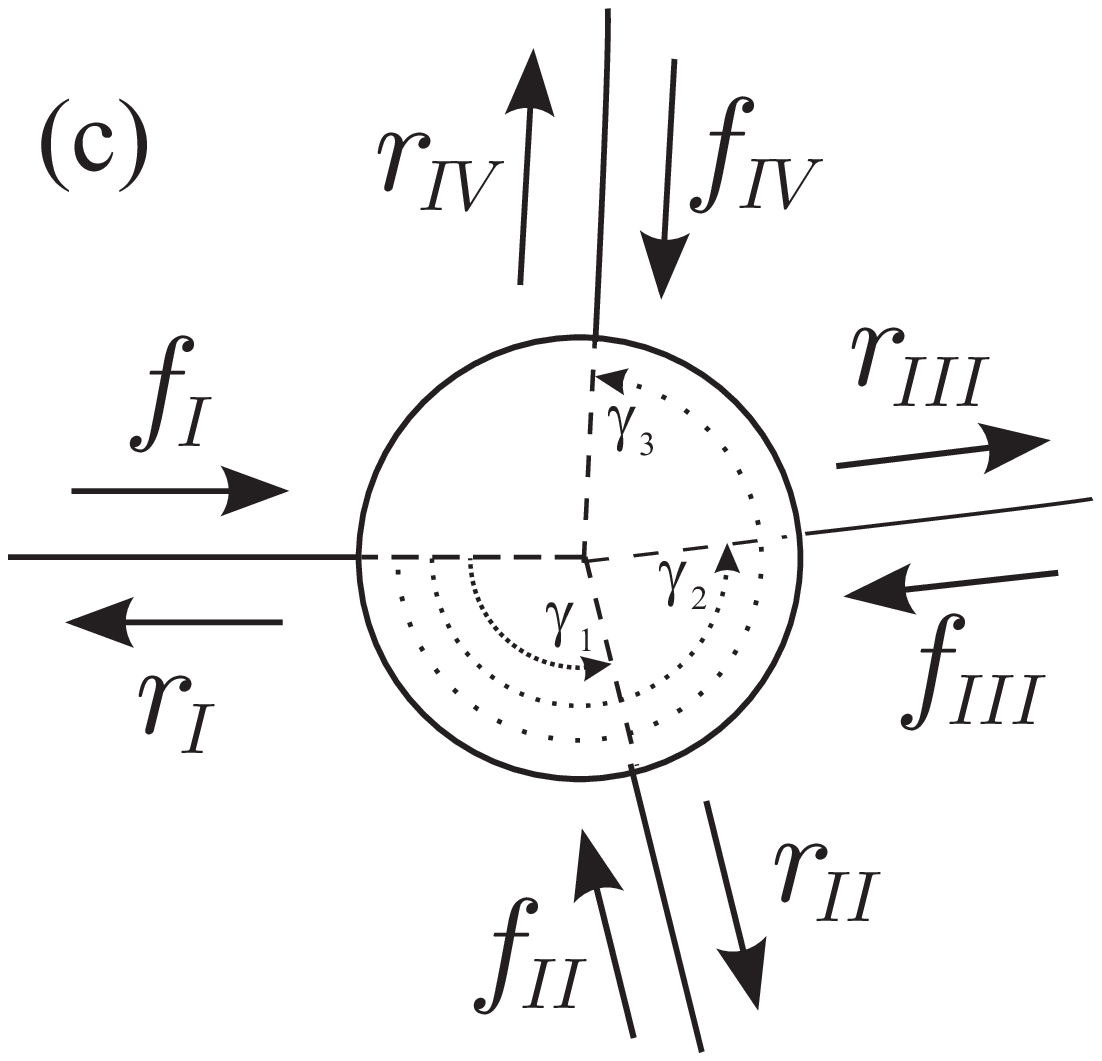}}\newline
\caption{The notations used for the spinor part of the wave functions in the
case of two- \subref{fig1a}, three- \subref{fig1b} and four-terminal rings 
\subref{fig1c}.}
\label{fig1}
\end{figure}
%%%%%%%%%%%%%%%%%%%%%%%%%%%%%%%%%%%%%%%%%%%%%%%%%%%%%%%%%%%%%%%%%%%%%%%%%%%%%

The building blocks of the rectangular arrays we investigate are two-,
three- and four-terminal quantum rings (see Fig.~\ref{fig1}), where, in
general, the boundary conditions allow both incoming and outgoing spinor
valued wave functions at each terminal: $\Psi
_{i}=f_{i}e^{ikx_{i}}+r_{i}e^{-ikx_{i}}$ ($i=I,I\!I,I\!I\!I,I\!V$), where $%
x_{i}$ denotes the local coordinate in terminal $i$. Note that the
amplitudes $f_{I},r_{I},f_{I\!I}$, etc. refer to two-component spinors, e.g. 
$f_{I}=%
\begin{pmatrix}
\left( f_{I}\right) _{\uparrow } \\ 
\left( f_{I}\right) _{\downarrow }%
\end{pmatrix}%
$. For the sake of definiteness, we focus on a general three-terminal ring,
shown in Fig.~\ref{fig1b}. The scattering problem in the case of a ring with
four terminals (Fig.~\ref{fig1c}) can also be solved analytically, as
presented in the Appendix, where we also give the results for a general
two-terminal ring (Fig.~\ref{fig1a}). The outgoing spinors ($r_{i}$, $%
i=I,I\!I,I\!I\!I$) are connected to the incoming ones ($f_{i})$ by $2\times
2 $ matrices, which can be determined by requiring the continuity of the
wave functions, and vanishing net spin current densities (Griffith
conditions) \cite{G53,X92,MPV04,FMBP05} at the junctions. For the same
boundary conditions as in Ref.~\onlinecite{KFBP08}, i.e. for $%
f_{I\!I},f_{I\!I\!I}=0$ in Fig.~\ref{fig1b}, the reflection matrix which
connects $r_{I}$ to the incoming spinor $f_{I}$ is given by%
\begin{eqnarray}
R_{_{\uparrow \uparrow }}^{f_{I}} &=&\varrho ^{\left( 1\right) }\cos
^{2}(\theta /2)+\varrho ^{\left( 2\right) }\sin ^{2}(\theta /2)-1,  \notag \\
R_{\uparrow \downarrow }^{f_{I}} &=&(\varrho ^{\left( 1\right) }-\varrho
^{\left( 2\right) })\sin (\theta /2)\cos (\theta /2),  \notag \\
R_{\downarrow \uparrow }^{f_{I}} &=&R_{\uparrow \downarrow }^{f_{I}},
\label{R_3_term} \\
R_{\downarrow \downarrow }^{f_{I}} &=&\varrho ^{\left( 1\right) }\sin
^{2}(\theta /2)+\varrho ^{\left( 2\right) }\cos ^{2}(\theta /2)-1,  \notag
\end{eqnarray}%
where 
\begin{eqnarray*}
\varrho ^{\left( \mu \right) }\! &=&\!8ka/y^{\left( \mu \right) }\left\{
-i(q^{\left( \mu \right) })^{2}\sin (2q^{\left( \mu \right) }\pi )\right. \\
&&\left. -kaq^{\left( \mu \right) }\left[ \sin (q^{\left( \mu \right)
}\gamma _{1})\sin (q^{\left( \mu \right) }(2\pi \!-\!\gamma _{1}))\right.
\right. \\
&&\left. \left. +\sin (q^{\left( \mu \right) }\gamma _{2})\sin (q^{\left(
\mu \right) }(2\pi \!-\!\gamma _{2}))\right] \right. \\
&&\left. +ik^{2}\!a^{2}\!\sin (q^{\left( \mu \right) }\gamma _{1})\sin
(q^{\left( \mu \right) }(\gamma _{2}\!-\!\gamma _{1}))\right. \\
&&\left. \times \sin (q^{\left( \mu \right) }(2\pi \!-\!\gamma
_{2}))\right\} ,
\end{eqnarray*}%
and 
\begin{eqnarray*}
y^{(\mu )}\! &=&\!8(q^{(\mu )})^{3}\!\left[ \cos \!\left[ (\left( -1\right)
^{\mu +1}\!w\!+\!2\phi )\pi \right] \!+\!\cos (2q^{\left( \mu \right) }\!\pi
)\!\right] \\
&&-12ika\!(q^{\left( \mu \right) })^{2}\!\sin (2q^{\left( \mu \right) }\pi
)+4k^{2}\!a^{2}\!q^{\left( \mu \right) }\!\cos (2q^{\left( \mu \right) }\pi )
\\
&&-2k^{2}\!a^{2}\!q^{\left( \mu \right) }\!\left[ \cos (2q^{\left( \mu
\right) }(\pi \!-\!\gamma _{2}\!+\!\gamma _{1}))\!-\!\cos (2q^{\left( \mu
\right) }\pi )\right. \\
&&\left. +\!\cos (2q^{\left( \mu \right) }(\pi \!-\!\gamma _{2}))\!+\!\cos
(2q^{\left( \mu \right) }(\pi \!-\!\gamma _{1}))\right] \\
&&+ik^{3}\!a^{3}\!\left[ \sin (2q^{\left( \mu \right) }(\pi \!-\!\gamma
_{2}\!+\!\gamma _{1}))\!-\!\sin (2q^{\left( \mu \right) }\pi )\right. \\
&&\left. +\!\sin (2q^{\left( \mu \right) }(\pi \!-\!\gamma _{1}))-\sin
(2q^{\left( \mu \right) }(\pi \!-\!\gamma _{2}))\right] ,
\end{eqnarray*}%
with $\phi =\Phi /\Phi _{0}$. The matrices describing the connection between
the outgoing spinors $r_{I\!I},$ $r_{I\!I\!I}$ and the input $f_{I}$ -- the
so called transmission matrices -- are given by 
\begin{eqnarray}
\left( T_{n}^{f_{I}}\right) _{\uparrow \uparrow }\! &=&\!e^{-i\gamma
_{n}/2}\left( \tau _{n}^{\left( 1\right) }\cos ^{2}(\theta /2)\!+\!\tau
_{n}^{\left( 2\right) }\sin ^{2}(\theta /2)\right) ,  \notag \\
\left( T_{n}^{f_{I}}\right) _{\uparrow \downarrow }\! &=&\!e^{-i\gamma
_{n}/2}\left( \tau _{n}^{\left( 1\right) }\!-\!\tau _{n}^{\left( 2\right)
}\right) \sin (\theta /2)\cos (\theta /2),  \notag \\
\left( T_{n}^{f_{I}}\right) _{\downarrow \uparrow }\! &=&\!e^{i\gamma
_{n}/2}\left( \tau _{n}^{\left( 1\right) }\!-\!\tau _{n}^{\left( 2\right)
}\right) \sin (\theta /2)\cos (\theta /2),  \label{T_3_term} \\
\left( T_{n}^{f_{I}}\right) _{\downarrow \downarrow }\! &=&\!e^{i\gamma
_{n}/2}\left( \tau _{n}^{\left( 1\right) }\sin ^{2}(\theta /2)\!+\!\tau
_{n}^{\left( 2\right) }\cos ^{2}(\theta /2)\right) ,  \notag
\end{eqnarray}%
where $n=1,2$, indicating the two possible output channels and 
\begin{eqnarray*}
\tau _{1}^{\left( \mu \right) } &=&\frac{8kaq^{\left( \mu \right) }}{%
y^{\left( \mu \right) }}e^{i\gamma _{1}/2\left( \left( -1\right) ^{\mu
+1}w+2\phi \right) }\times \\
&&\left\{ -ka\sin (q^{\left( \mu \right) }(\gamma _{2}-\gamma _{1}))\sin
(q^{\left( \mu \right) }(2\pi -\gamma _{2}))\right. \\
&&\left. +iq^{\left( \mu \right) }\left[ e^{-i\pi \left( \left( -1\right)
^{\mu +1}w+2\phi \right) }\sin (q^{\left( \mu \right) }\gamma _{1})\right.
\right. \\
&&\left. \left. -\sin (q^{\left( \mu \right) }(2\pi -\gamma _{1}))\right]
\right\} , \\
\tau _{2}^{\left( \mu \right) } &=&\frac{8kaq^{\left( \mu \right) }}{%
y^{\left( \mu \right) }}e^{i\gamma _{2}/2\left( \left( -1\right) ^{\mu
+1}w+2\phi \right) }\times \\
&&\left\{ kae^{-i\pi \left( \left( -1\right) ^{\mu +1}w\!+\!2\phi \right)
}\!\sin (q^{\left( \mu \right) }\!\gamma _{1})\sin (q^{\left( \mu \right)
}\!(\gamma _{2}\!-\!\gamma _{1}))\right. \\
&&\left. +iq^{\left( \mu \right) }\left[ e^{-i\pi \left( \left( -1\right)
^{\mu +1}w+2\phi \right) }\sin (q^{\left( \mu \right) }\gamma _{2})\right.
\right. \\
&&\left. \left. -\sin (q^{\left( \mu \right) }(2\pi -\gamma _{2}))\right]
\right\} .
\end{eqnarray*}%
Note that the boundary conditions applied to obtain the $R^{f_{I}}$ and $%
T_{n}^{f_{I}}$ matrices above are similar to that of Ref.~\onlinecite{KFBP08}%
. However the magnetic field induced shift of the spin Zeeman levels leads
to a doubling of the parameters according to Eq.~(\ref{q_mu}). This modifies
significantly the physical transport properties of the device.

Let us point out that having obtained the matrix elements above is enough to
handle the problem with both incoming and outgoing waves on all terminals of
the ring as shown in Fig.~\ref{fig1b}. Namely, we can consider the three
inputs $f_{i}$ ($i=I,I\!I,I\!I\!I$) separately and determine the
corresponding reflection and transmission matrices. The outputs in the
superposed problem will consist of contributions from all inputs: the
reflected part of the spinor which enters on the same lead, and the
transmitted parts of the other two inputs into the respective lead: 
\begin{eqnarray}
r_{I}
&=&R^{f_{I}}f_{I}+T_{2}^{f_{I\!I}}f_{I\!I}+T_{1}^{f_{I\!I\!I}}f_{I\!I\!I}, 
\notag \\
r_{I\!I}
&=&T_{1}^{f_{I}}f_{I}+R^{f_{I\!I}}f_{I\!I}+T_{2}^{f_{I\!I\!I}}f_{I\!I\!I},
\label{rI_rII_rIII} \\
r_{I\!I\!I}
&=&T_{2}^{f_{I}}f_{I}+T_{1}^{f_{I\!I}}f_{I\!I}+R^{f_{I\!I\!I}}f_{I\!I\!I}. 
\notag
\end{eqnarray}%
Considering $f_{I\!I}$ ($f_{I\!I\!I}$) as the only input, the reflection and
transmission matrices are the same as those for the input $f_{I}$, except
for the appropriate changes of the angles, since in the reference frame of $%
f_{I\!I}$ ($f_{I\!I\!I}$), the angles of the output leads are measured from
the lead through which $f_{I\!I}$ ($f_{I\!I\!I}$) enters the ring. In order
to get the contributions to the output spinors for the input $f_{I\!I}$ ($%
f_{I\!I\!I}$) in the reference frame of $f_{I}$, the matrices need to be
rotated (see Fig.~\ref{fig1b}) by the angle of $\gamma _{1}$($\gamma _{2}$): 
\begin{equation}
M^{f_{I\!I}}=U_{\gamma _{1}}M_{\substack{ \gamma _{1}\leftrightarrow \gamma
_{2}-\gamma _{1}  \\ \gamma _{2}\leftrightarrow 2\pi -\gamma _{1}}}%
^{f_{I}}U_{\gamma _{1}}^{-1}  \label{3_Ru_Tu}
\end{equation}%
\begin{equation}
M^{f_{I\!I\!I}}=U_{\gamma _{2}}M_{\substack{ \gamma _{1}\leftrightarrow 2\pi
-\gamma _{2}  \\ \gamma _{2}\leftrightarrow 2\pi -\gamma _{2}+\gamma _{1}}}%
^{f_{I}}U_{\gamma _{2}}^{-1}  \label{3_Rv_Tv}
\end{equation}%
where $M=R,T_{1},T_{2}$ and 
\begin{equation*}
U_{\gamma _{n}}=%
\begin{pmatrix}
e^{-i\frac{\gamma _{n}}{2}} & 0 \\ 
0 & e^{i\frac{\gamma _{n}}{2}}%
\end{pmatrix}%
,\quad n=1,2.
\end{equation*}

The above approach is also valid in the case of the two- and four-terminal
rings. Using the reflection and transmission matrices as presented in the
Appendix, the more general problem of having both incoming and outgoing
waves on all terminals can easily be treated. All possible reflections and
transmissions can thus be taken into account when forming two-dimensional
arrays of such rings.

\section{Rectangular arrays of quantum rings}

\label{arraysec}

\subsection{Magnetoconductance properties}

\label{arraysecsubsec1}

\begin{figure}[tbh]
\begin{center}
\includegraphics*[width=8.0cm]{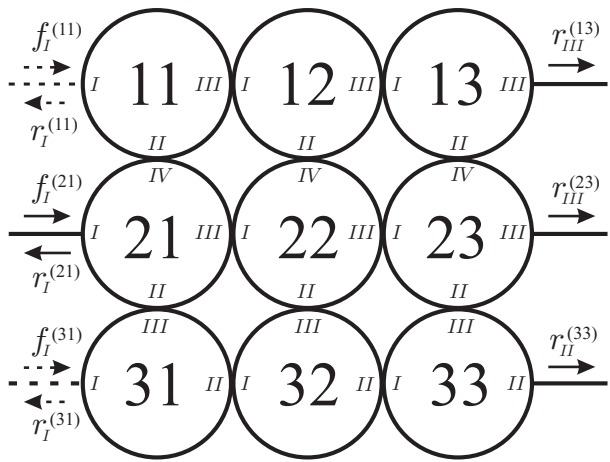} \vspace{%
-0.4cm}
\end{center}
\caption{ The geometry of the device in the simplest case of a 3$\times$3
array with three or one (without leads displayed with dashed lines) input
terminals. The notations can easily be generalized to larger arrays.}
\label{networkfig}
\end{figure}

Based on the analytic results presented in the previous section and in the
Appendix we may build N$\times $M two-dimensional rectangular arrays of
quantum rings, where both perpendicular electric and magnetic fields are
present, so that the former one can be used to change the strength of the
SOI.~\cite{NATE97} Here we focus on of 3$\times $3,\ 4$\times $4, and 5$%
\times $5\ arrays and assume that neighboring rings touch each other. In
addition, we limit ourselves to arrays that are closed in the vertical, and
open in the horizontal direction, as shown in Fig.~\ref{networkfig}. Two
types of such arrays will be investigated: i) the electron can enter/exit
the array through any of the rings in the horizontal direction, ii) the
electron can enter the array through one ring only (no leads are attached to
the other rings on the entrance side), but can exit through any of the rings
on the opposite side (Fig.~\ref{networkfig} without the dashed curves). In
both cases the conductance is derived from the linear set of equations
resulting from the fit of the wave functions $\Psi _{i}^{\left( kl\right) }$
($i=I,I\!I,I\!I\!I,I\!V$ and \thinspace $k,l=1..N$, where $N$ is the number
of rings along one direction in the array) and their derivatives $\partial
_{x_{i}^{\left( kl\right) }}\Psi _{i}^{\left( kl\right) }$ in the points,
where the rings touch each other, i.e. for example:%
\begin{eqnarray}
\left. \Psi _{I\!I\!I}^{\left( 11\right) }\right\vert _{x_{I\!I\!I}^{\left(
11\right) }=0} &=&\left. \Psi _{I}^{\left( 12\right) }\right\vert
_{x_{I}^{\left( 12\right) }=0},  \notag \\
\left. \partial _{x_{I\!I\!I}^{\left( 11\right) }}\Psi _{I\!I\!I}^{\left(
11\right) }\right\vert _{x_{I\!I\!I}^{\left( 11\right) }=0} &=&-\left.
\partial _{x_{I}^{\left( 12\right) }}\Psi _{I}^{\left( 12\right)
}\right\vert _{x_{I}^{\left( 12\right) }=0}.  \label{fit}
\end{eqnarray}%
Here we used the notations of Fig.~\ref{networkfig}. (Note that the negative
sign in Eq. (\ref{fit}) is a consequence of the opposite direction of the
local coordinates in the leads $III$ of ring $\left\{ 11\right\} $ and $I$
of ring $\left\{12\right\}$) \ Eqs. (\ref{fit}) lead to%
\begin{eqnarray*}
f_{I\!I\!I}^{\left( 11\right) }+r_{I\!I\!I}^{\left( 11\right) }
&=&f_{I}^{\left( 12\right) }+r_{I}^{\left( 12\right) }, \\
f_{I\!I\!I}^{\left( 11\right) }-r_{I\!I\!I}^{\left( 11\right) }
&=&-f_{I}^{\left( 12\right) }+r_{I}^{\left( 12\right) },
\end{eqnarray*}%
from which follows that 
\begin{eqnarray*}
f_{I\!I\!I}^{\left( 11\right) } &=&r_{I}^{\left( 12\right) }, \\
r_{I\!I\!I}^{\left( 11\right) } &=&f_{I}^{\left( 12\right) },
\end{eqnarray*}%
i.e., the spinor entering (exiting) ring $\left\{11\right\}$ on terminal $%
I\!I\!I$ is equal to the spinor exiting (entering) ring $\left\{12\right\}$
on terminal $I$. The spinors $r_{I\!I\!I}^{\left( 11\right) }$ and $%
r_{I}^{\left( 12\right) }$ can be given with the help of the reflection and
transmission matrices of a three-terminal ring according to Eqs. (\ref%
{rI_rII_rIII}).\ 

For a small number of rings the resulting set of equations can be solved
analytically, however already for an array of 3$\times $3 rings shown in
Fig.~\ref{networkfig}, it consists of 60 equations, which is preferably
solved by numerical means, although analytic solutions exist in principle.
(For larger arrays the number of equations scales practically with the
number of rings). After having determined the output spinor valued wave
functions $r_{I\!I\!I}^{\left( 1N\right) },r_{I\!I\!I}^{\left( 2N\right)
},...r_{I\!I}^{\left( NN\right) }$, where $N$ is the number of rings in the
horizontal direction, the Landauer--B\"{u}ttiker \cite{D95} formula 
\begin{equation*}
G=G_{\uparrow }+G_{\downarrow }
\end{equation*}%
where 
\begin{eqnarray*}
G_{\uparrow }\! &=&\!\frac{e^{2}}{h}\left( \left\vert (r_{I\!I\!I}^{\left(
1N\right) })_{\uparrow }\right\vert ^{2}\!+\!\left\vert (r_{I\!I\!I}^{\left(
2N\right) })_{\uparrow }\right\vert ^{2}\!+...+\!\left\vert
(r_{I\!I}^{\left( NN\right) })_{\uparrow }\right\vert ^{2}\right) , \\
G_{\downarrow }\! &=&\!\frac{e^{2}}{h}\left( \left\vert (r_{I\!I\!I}^{\left(
1N\right) })_{\downarrow }\right\vert ^{2}\!+\!\left\vert
(r_{I\!I\!I}^{\left( 2N\right) })_{\downarrow }\right\vert
^{2}\!+...+\!\left\vert (r_{I\!I}^{\left( NN\right) })_{\downarrow
}\right\vert ^{2}\right) ,
\end{eqnarray*}%
is used to calculate the conductance of the arrays, averaged over the two $%
\sigma _{z}$ eigenspinor inputs. We note that our method of using single
rings as building blocks can easily be used to determine the conductance of
arrays of arbitrary -- not necessarily rectangular -- configuration as well.

%%%%%%%%%%%%%%%%%%%%%%%%%%%%%%%%%%%%%%%%%%%%%%%%%%%%%%%%%%%%%%%%%%%%%%%%%%%%%%
%%%%%%%%%%%%%%%%%%%%%%%%%%%%%%%%%%%%%%%%%%%%%%%%%%%%%%%%%%%%%%%%%%%%%%%%%%%%%%

\begin{figure}[tbh]
\centering
\subfigure{
\label{fig3a}
\includegraphics[width=8cm]{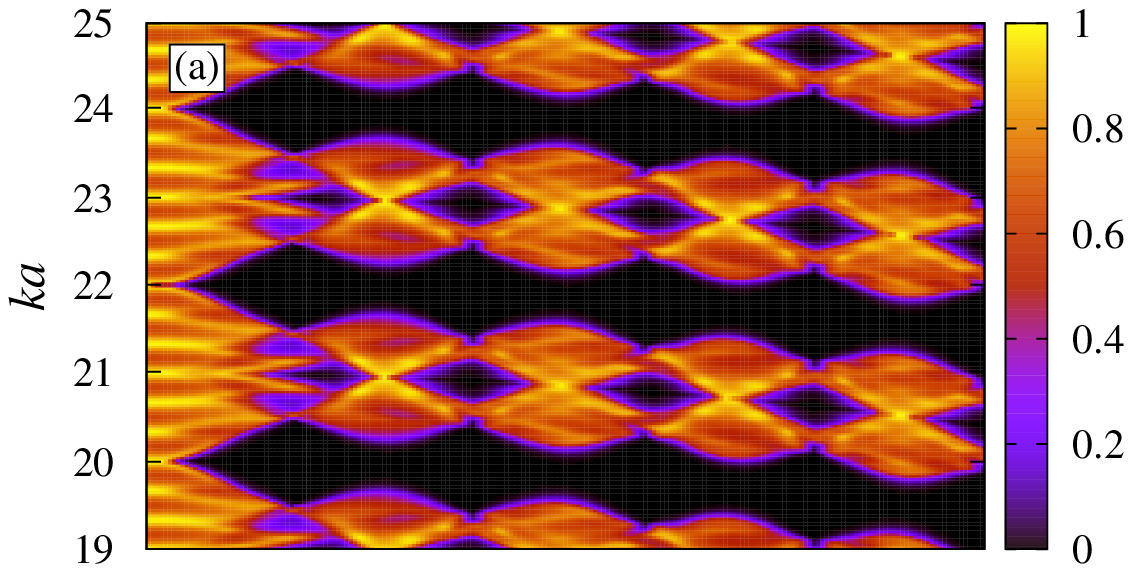}}\newline
\subfigure{
\label{fig3b}
\includegraphics[width=8cm]{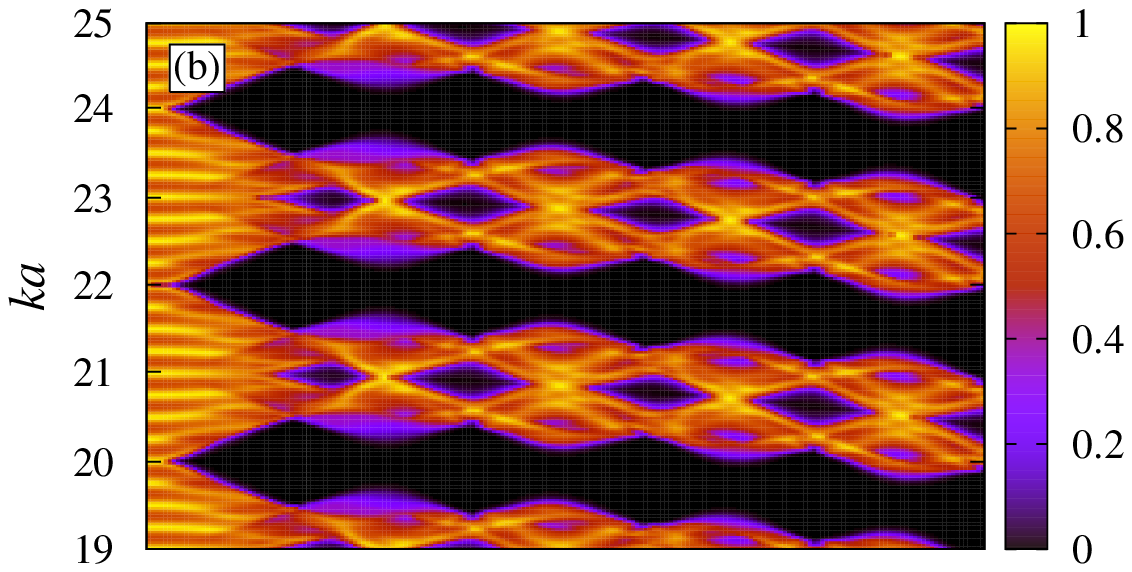}}\newline
\subfigure{
\label{fig3c}
\includegraphics[width=8cm]{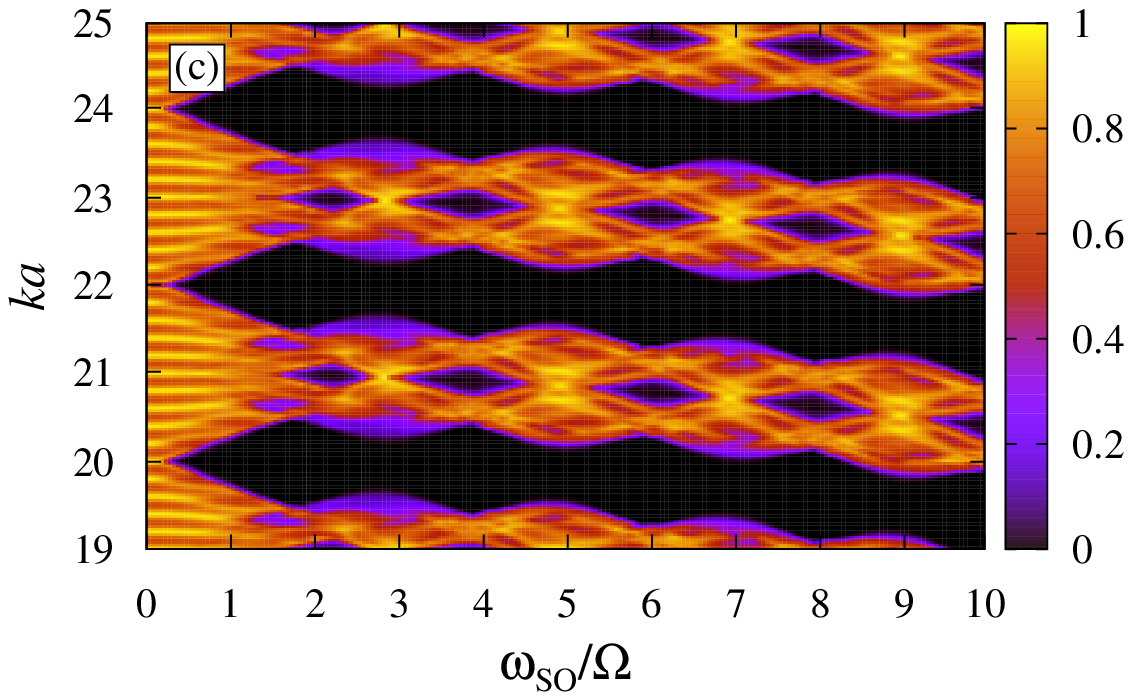}}\newline
\caption{(color online) The conductance $G/G_{0}$ (%
$G_{0}=e^2/h$) of a 3$\times$3 \subref{fig3a}, 4$\times$4 
\subref{fig3b} and 5$\times$5 \subref{fig3c} rectangular array with 3, 4,
and 5 input terminals, respectively, for zero magnetic flux as a function of
the SOI strength and $ka$.}
\label{conductancefig5}
\end{figure}

Figure \ref{conductancefig5} shows a contour plot of the conductance (in $%
e^{2}/h$ units)\ of rectangular arrays of 3$\times $3, 4$\times $4 and 5$%
\times $5 quantum rings, for zero magnetic flux as a function of the SOI
strength $\omega _{\text{SO}}/\Omega $ and $ka$. The values of $ka$ are
varied around $k_{F}a=20.4$, corresponding to a Fermi energy $11.13$ meV in
case of an effective mass $m^{\ast }=0.023m$ of InAs and rings of radius $%
a=0.25$ $\mu $m. In two-dimensional electron systems within an InAs quantum
well, the value of $\alpha $ can be varied \cite{G00,NATE97}\ up to $40$
peVm. The different
arrays show similar behavior for larger values of the SOI strength: there
are slightly downwards bending stripes (initially around even values of $ka$%
) where the devices are completely opaque for the electrons, and also
conducting regions which are initially around odd values of $ka$ and have
complex internal structure. Comparing our results to the case of a single
ring with diametrically coupled leads \cite{MPV04}, it can be seen that the
overall periodicity as a function of $ka$ is determined by single-ring
interferences. The increasing number of the rings causes modulations
superimposing on the single-ring behavior. This point is probably the most
apparent if we recall \cite{MPV04} that zero conductance areas are simply
lines on the $ka-\omega _{\text{SO}}/\Omega $ plane for a single
two-terminal ring, while in our case there are stripes, the width of which
is slightly increasing with the size of the array. This effect is related to
the increasing number of consecutive partially destructive interferences
that finally lead to essentially zero currents at the outputs. Additionally,
if we considered an infinite network, the periodic boundary conditions would
allow only discrete values of $ka$ for a given SOI strength with nonzero
conductance. Thus the results presented in Fig.~\ref{conductancefig5}
demonstrate a transition between the conductance properties of a single ring
and that of an infinite network.

%%%%%%%%%%%%%%%%%%%%%%%%%%%%%%%%%%%%%%%%%%%%%%%%%%%%%%%%%%%%%%%%%%%%%%%%%%%%%%

%%%%%%%%%%%%%%%%%%%%%%%%%%%%%%%%%%%%%%%%%%%%%%%%%%%%%%%%%%%%%%%%%%%%%%%%%%%%%%

Focusing on small values of $\omega _{\text{SO}}/\Omega $, Fig.~\ref%
{conductancefig5} shows a narrowing of the non-conducting regions until they
eventually disappear when no SOI is present. Here the conductance still
depends on $ka$, but its minimal values are not zeros and a periodic
behavior can be seen: for a network of $N\times N$ rings, there are $N$
minima as the value of $ka$ is increased by $1.$ This size-dependent
modulation is related to the horizontal extent of the device: If we compare
the conductance of the networks to that of rings of the same size and number
without vertical connections, the same periodic behavior can be seen around
zero SOI.

%%%%%%%%%%%%%%%%%%%%%%%%%%%%%%%%%%%%%%%%%%%%%%%%%%%%%%%%%%%%%%%%%%%%%%%%%%%%%%
\begin{figure}[tbh]
\centering
\subfigure{
\label{fig4a}
\includegraphics[width=8cm]{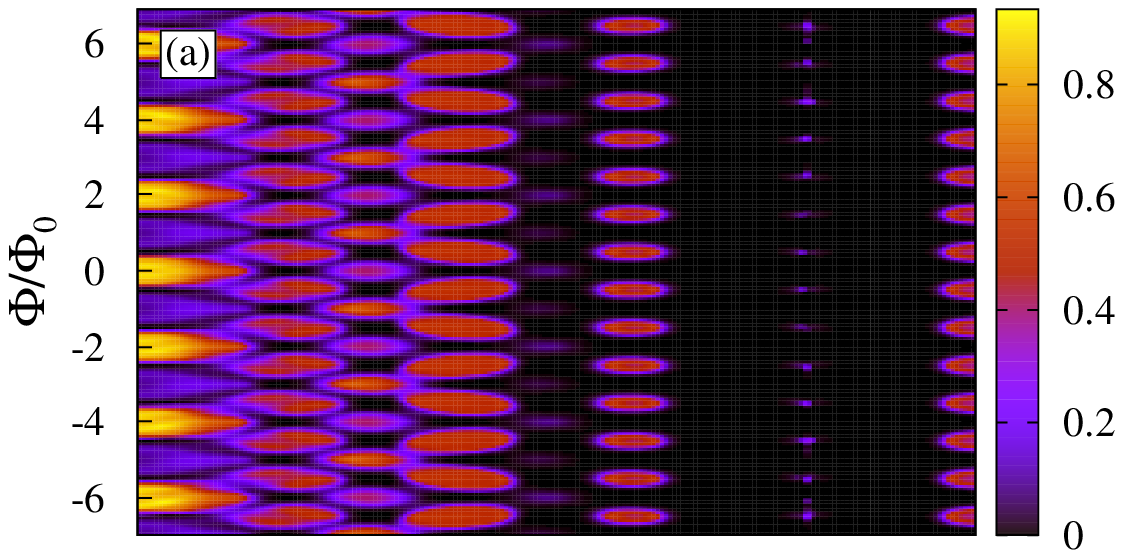}}\newline
\subfigure{
\label{fig4b}
\includegraphics[width=8cm]{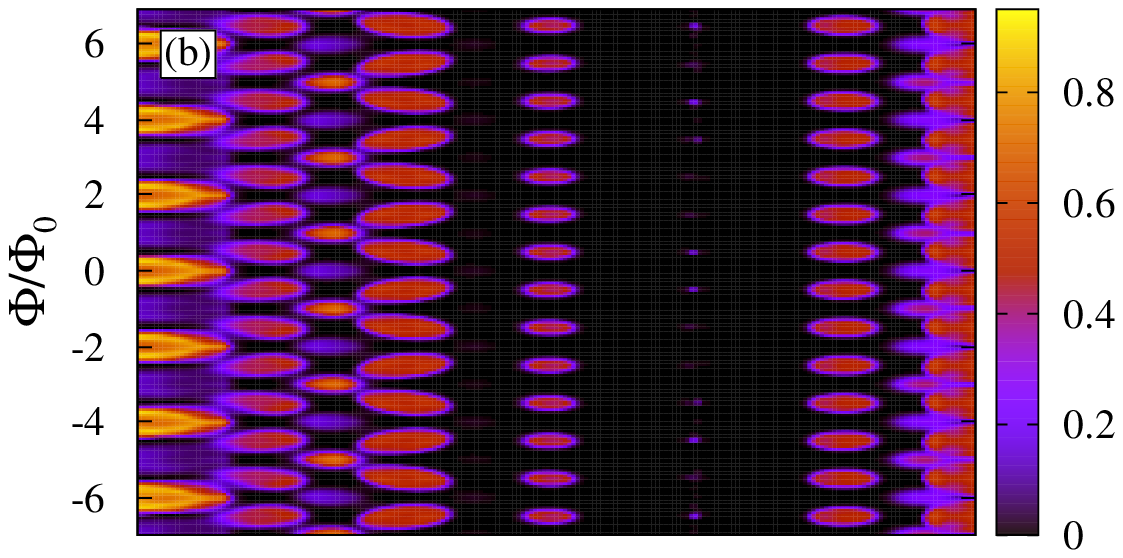}}\newline
\subfigure{
\label{fig4c}
\includegraphics[width=8cm]{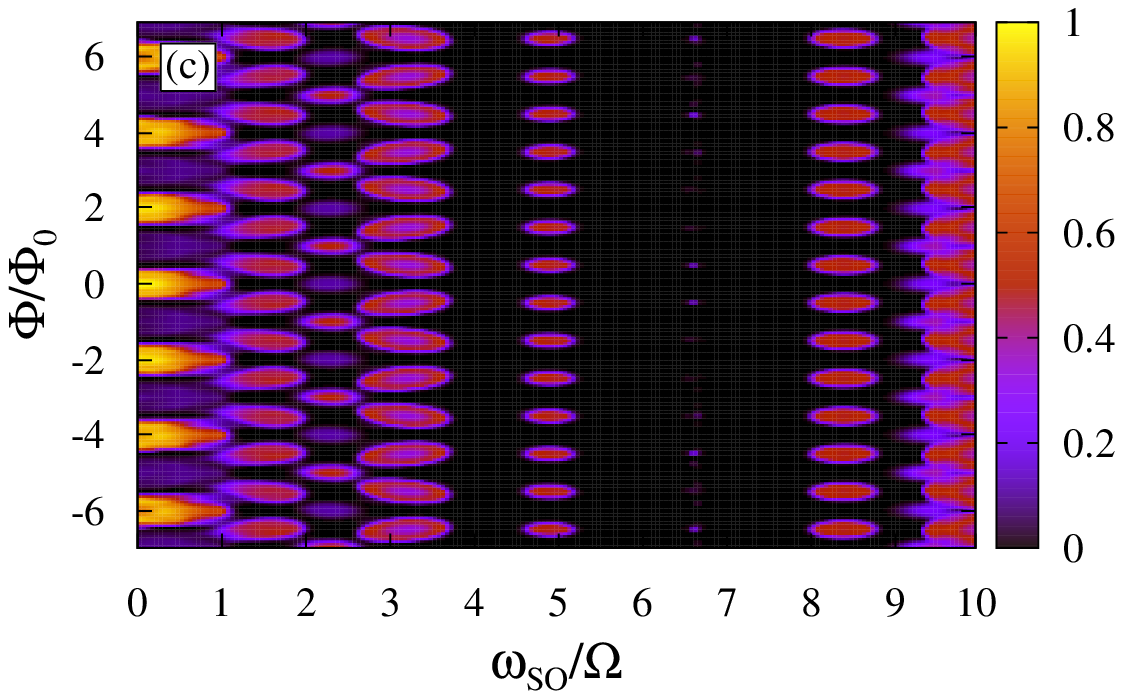}}\newline
\caption{(color online) The conductance $G/G_{0}$ (%
$G_{0}=e^2/h$) of a 3$\times$3 \subref{fig4a}, 4$\times$4 
\subref{fig4b} and 5$\times$5 \subref{fig4c} rectangular array with 3, 4,
and 5 input terminals, respectively, for $ka=19.6$ as a function of the SOI
strength and the magnetic flux $\Phi $ (in units of $\Phi_{0}=h/e$).}
\label{magnetoconductancefig5}
\end{figure}
%%%%%%%%%%%%%%%%%%%%%%%%%%%%%%%%%%%%%%%%%%%%%%%%%%%%%%%%%%%%%%%%%%%%%%%%%%%%%%

Figure~\ref{magnetoconductancefig5} shows the normalized magnetoconductance
of networks of 3$\times $3, 4$\times $4 and 5$\times $5 quantum rings for $%
ka=19.6$ as a function of the SOI strength and the magnetic flux $\Phi $
(measured in units of $\Phi _{0}$). When $\omega _{\text{SO}}/\Omega $ is
zero, Aharonov-Bohm (AB) oscillations appear. For larger values of $\omega _{%
\text{SO}}/\Omega $ both AB and Aharonov-Casher \cite{AC84} oscillations can
be seen in the magnetoconductance. As Fig.~\ref{magnetoconductancefig5} was
plotted for a certain value of $ka,$ the effect of the bending
non-conducting stripes seen in Fig.~\ref{conductancefig5} can also be seen
as the decrease of the conductance when such a stripe is reached due to the
change of the SOI strength, and its increase again, when the stripe is left.
We note that for larger values of $ka$ this bending effect is less
pronounced.

%%%%%%%%%%%%%%%%%%%%%%%%%%%%%%%%%%%%%%%%%%%%%%%%%%%%%%%%%%%%%%%%%%%%%%%%%%%%%%
\begin{figure}[tbh]
\centering
\includegraphics[width=8cm]{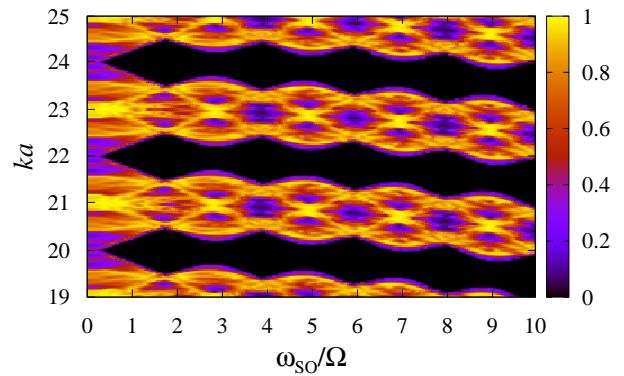}\newline
\caption{(color online) The conductance $G/G_{0}$ (%
$G_{0}=e^2/h$) of a 5$\times$5 rectangular array with a single
input lead attached to ring $\left\{31\right\}$, for zero magnetic flux as a
function of the SOI strength and $ka$.}
\label{conductancefig1}
\end{figure}
%%%%%%%%%%%%%%%%%%%%%%%%%%%%%%%%%%%%%%%%%%%%%%%%%%%%%%%%%%%%%%%%%%%%%%%%%%%%%%

%%%%%%%%%%%%%%%%%%%%%%%%%%%%%%%%%%%%%%%%%%%%%%%%%%%%%%%%%%%%%%%%%%%%%%%%%%%%%%
\begin{figure}[tbh]
\centering
\includegraphics[width=8cm]{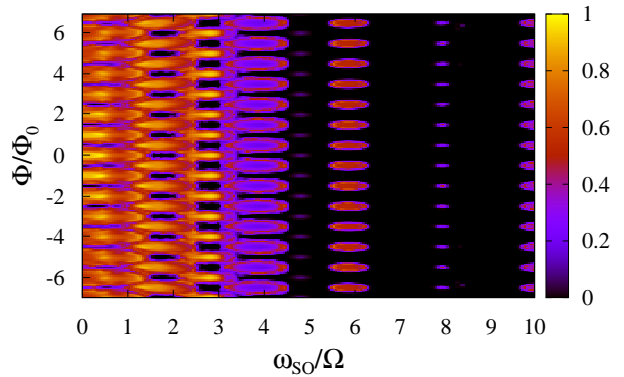}%
\newline
\caption{(color online) The conductance $G/G_{0}$ (%
$G_{0}=e^2/h$) of a 5$\times$5 rectangular array with a single
input lead attached to ring $\left\{31\right\}$, for $ka=19.57$ as a
function of the SOI strength and the magnetic flux $\Phi$ (in units of $%
\Phi_{0}=h/e $).}
\label{magnetoconductancefig1}
\end{figure}
%%%%%%%%%%%%%%%%%%%%%%%%%%%%%%%%%%%%%%%%%%%%%%%%%%%%%%%%%%%%%%%%%%%%%%%%%%%%%%

Figures \ref{conductancefig1} and \ref{magnetoconductancefig1} show the
conductance of a 5$\times $5 network with a single input lead in the middle
(i.e. attached to ring $\left\{31\right\}$, using the notations of Fig.~\ref%
{networkfig}) as a function of $ka,$ and $\omega _{\text{SO}}/\Omega $ (Fig.~%
\ref{conductancefig1}), and the magnetic field and $\omega _{\text{SO}%
}/\Omega $ (Fig.~\ref{magnetoconductancefig1}). The overall structure of
these plots remains the same as in the case when the current can enter
through all the rings on the left hand side, but the different boundary
conditions modify the fine structure of the plots.

\begin{figure}[tbh]
\centering
\subfigure{
\label{fig7a}
\includegraphics[width=8cm]{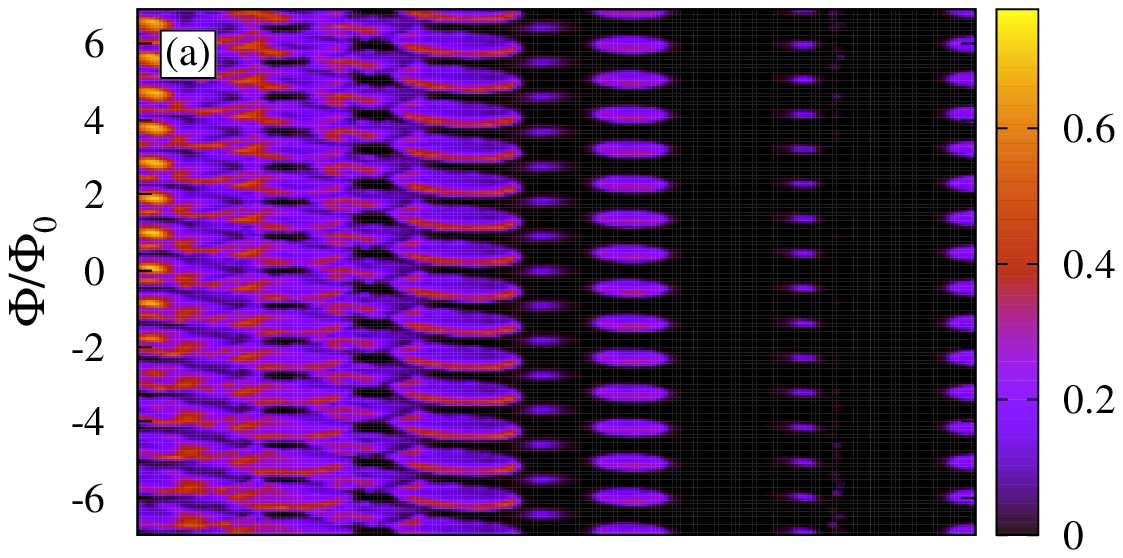}}%
\newline
\subfigure{
\label{fig7b}
\includegraphics[width=8cm]{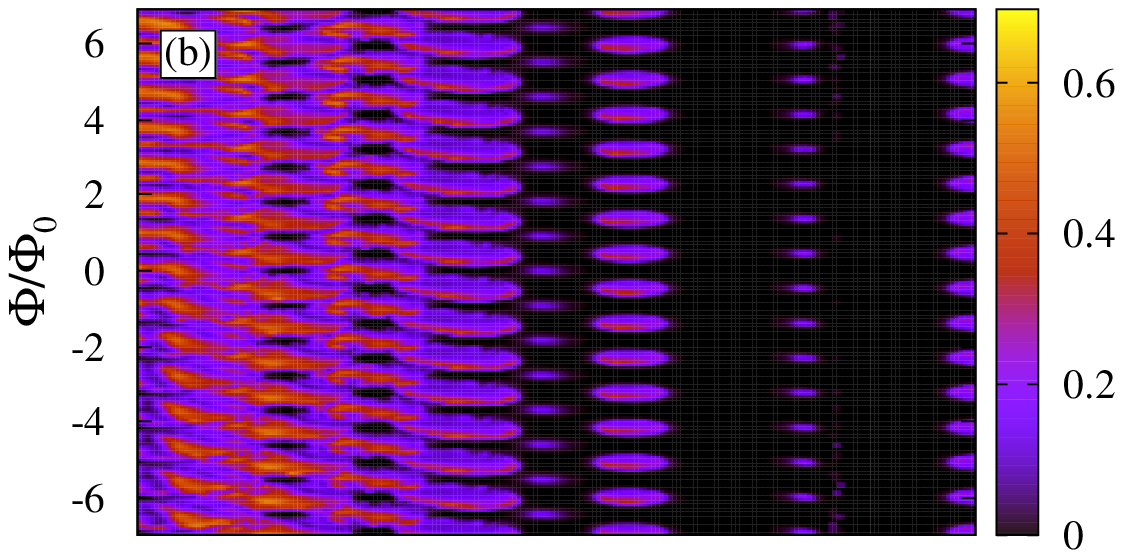}}%
\newline
\subfigure{
\label{fig7c}
\includegraphics[width=8cm]{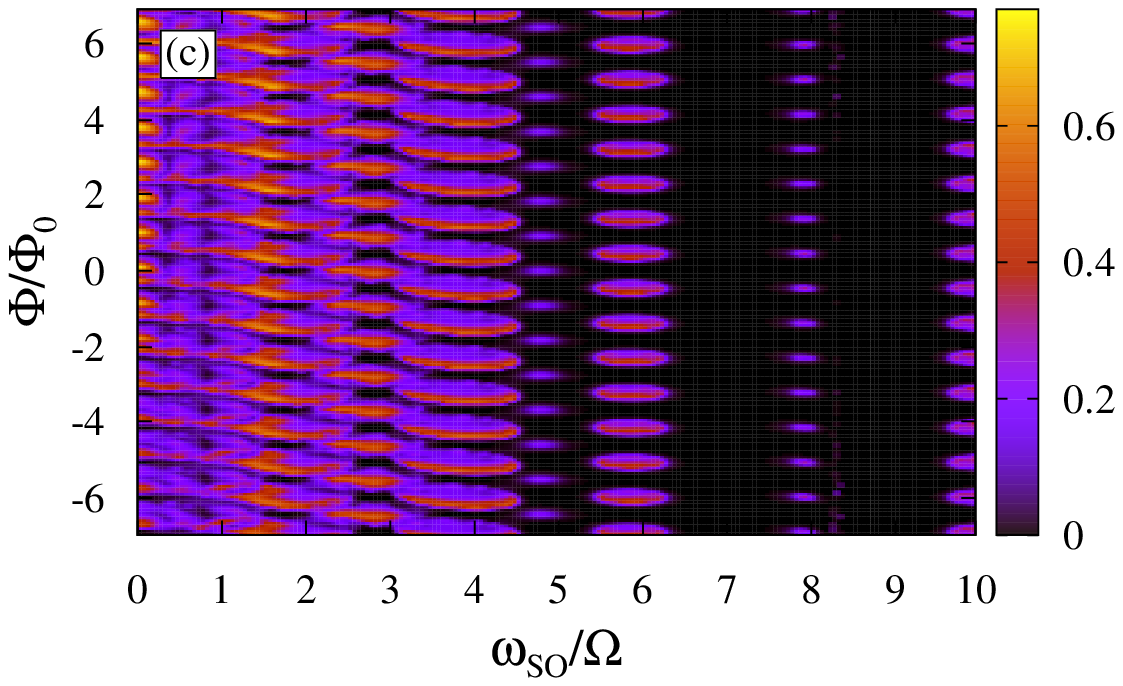}}%
\newline
\caption{(color online) The probabilities of the $|\uparrow_x\rangle$ 
\subref{fig7a}, $|\uparrow_y\rangle$ \subref{fig7b}, and $|\uparrow_z\rangle$ 
\subref{fig7c} outputs at ring $\left\{55\right\}$ of a 5$\times$5
rectangular array with one input lead (attached to ring $\left\{31\right\}$%
), for $ka=19.6$ as a function of the SOI strength and the magnetic flux $%
\Phi $ (in units of $\Phi_{0}=h/e$). The incoming spin state is chosen to be 
$|\uparrow_z\rangle$.}
\label{probabilityfig}
\end{figure}

\begin{figure}[tbh]
\centering
\includegraphics[width=8cm]{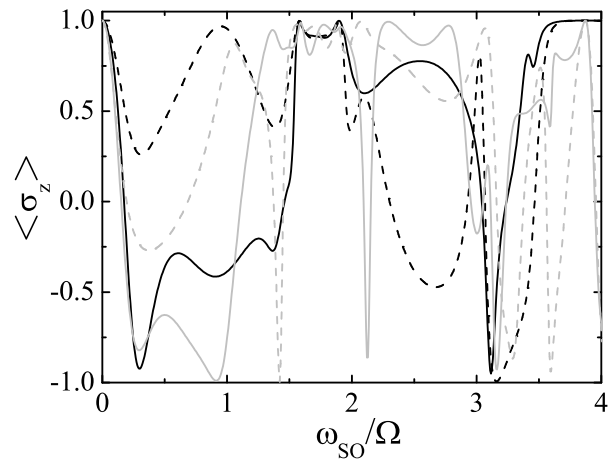}\newline
\caption{The spin transformation properties of a 5$\times$5 array with
input leads attached to all rings and only to ring $\left\{31\right\}$
(black and grey curves, respectively). The $z$ component of the normalized 
spin states transmitted via the output terminals attached to ring $\left\{25\right\}$
(solid line) and ring $\left\{45\right\}$ (dashed line). The incoming spin 
state is chosen to be $|\uparrow_z\rangle$.}
\label{rotfiguniform}
\end{figure}

Our method allows the calculation of the spin directions for the different
output terminals, and we found that spin-dependent interference in the array
results in nontrivial spin transformations. Fig.~\ref{probabilityfig} shows
the spin resolved transmission probabilities for a 5$\times $5 ring array
with a single input lead. The incoming spin state is chosen to be $|\uparrow
_{z}\rangle ,$ i.e., the spin-up eigenstate of $\sigma _{z},$ and the
contour plots show the probabilities of the $|\uparrow _{x}\rangle ,$ $%
|\uparrow _{y}\rangle $ and $|\uparrow _{z}\rangle $ outputs at ring $%
\left\{55\right\}$ on the right hand side. The fact that the $|\uparrow
_{z}\rangle $ input spinor changes its direction (as it is seen in Fig.~\ref%
{probabilityfig}, it can be transformed into $|\uparrow _{x}\rangle $ or $%
|\uparrow _{y}\rangle $) is due to the SOI induced spin rotations. The
actual values of the spin resolved transmission probabilities are determined
by the spin dependent interference phenomena. Fig.~\ref{rotfiguniform} shows
the $z$ component of the normalized output spinors and visualizes that spin resolved
results depend on the input side geometry as well. As we can see, the spin
components change in the whole available range between -1 and 1, and their
behavior is rather different for the cases when the electron can enter the
array through any of the five terminals, or only through the one attached to
ring $\left\{31\right\}$. This phenomenon together with other spin dependent
interference effects \cite{PC07,CCZ04,KNV04,ID03,ZS07,CAF06,BO08} can lead to spin
sensitive quantum networks.

%%%%%%%%%%%%%%%%%%%%%%%%%%%%%%%%%%%%%%%%%%%%%%%%%%%%%%%%%%%%%%%%%%%%%%%%%%%%%%
%%%%%%%%%%%%%%%%%%%%%%%%%%%%%%%%%%%%%%%%%%%%%%%%%%%%%%%%%%%%%%%%%%%%%%%%%%%%%%

%%%%%%%%%%%%%%%%%%%%%%%%%%%%%%%%%%%%%%%%%%%%%%%%%%%%%%%%%%%%%%%%%%%%%%%%%%%%%%
%%%%%%%%%%%%%%%%%%%%%%%%%%%%%%%%%%%%%%%%%%%%%%%%%%%%%%%%%%%%%%%%%%%%%%%%%%%%%%

\subsection{Effect of point-like scatterers}

\label{arraysecsubsec2}

Now we will investigate to what extent the conductance properties are
modified by the presence of random scatterers. Although high mobility
samples have already become available (such that at cryogenic temperatures
transport is found to be ballistic over tens of microns), considering also
the effects caused by scattering events provides a more realistic
description for most cases. To this end we introduce point-like scattering
centers between the rings. In particular, at each point $j$ where two rings
touch each other, we consider an additional Dirac delta potential of the
form $\eta _{j}\delta \left( j\right) $. Here $\eta _{j}$ represent
independent normally distributed random variables with zero mean, and
root-mean-square deviation $D$. By tuning $D$ we can model weak disturbances
(small $D$) as well as the case when frequent scattering events completely
change the character of the transport process (corresponding to large values
of $D$).

%%%%%%%%%%%%%%%%%%%%%%%%%%%%%%%%%%%%%%%%%%%%%%%%%%%%%%%%%%%%%%%%%%%%%%%%%%%%%%
\begin{figure}[tbh]
\centering
\includegraphics[width=8cm]{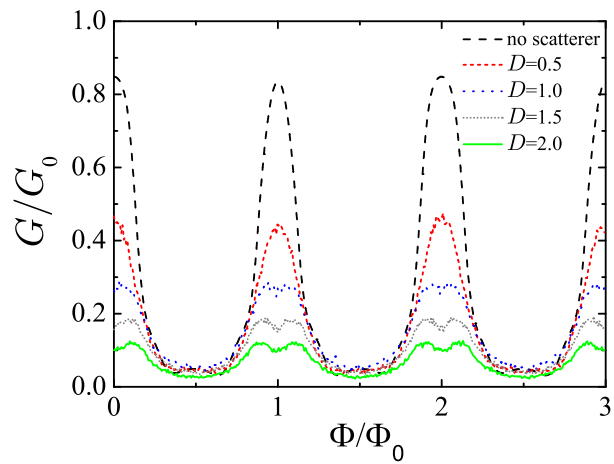}\newline
\caption{(color online) The conductance $G$ (in units of $%
G_{0}=e^2/h$) of a 5$\times$5 rectangular array with point-like
random scatterers between the rings for different root-mean-square
deviations $D$ as a function of the magnetic flux $\Phi$ (in units of $%
\Phi_{0}=h/e$) for $ka=20.2$ and $\protect\omega _{\text{SO}}/\Omega=13.0$.}
\label{scattfig1}
\end{figure}
%%%%%%%%%%%%%%%%%%%%%%%%%%%%%%%%%%%%%%%%%%%%%%%%%%%%%%%%%%%%%%%%%%%%%%%%%%%%%%

%%%%%%%%%%%%%%%%%%%%%%%%%%%%%%%%%%%%%%%%%%%%%%%%%%%%%%%%%%%%%%%%%%%%%%%%%%%%%%
\begin{figure}[tbh]
\centering
\includegraphics[width=8cm]{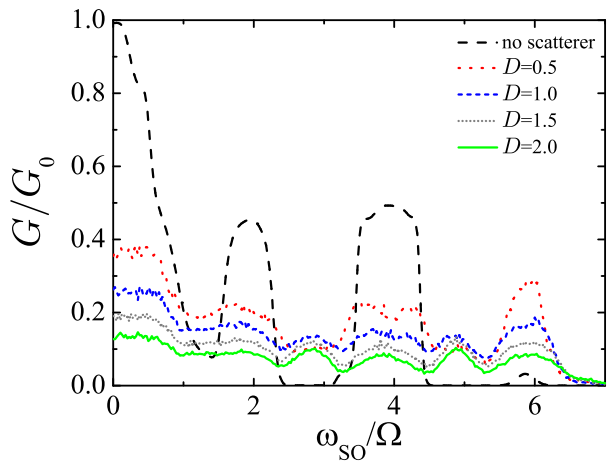}\newline
\caption{(color online) The conductance $G$ (in units of $%
G_{0}=e^2/h$) of a 5$\times$5 rectangular array with point-like
random scatterers between the rings for different root-mean-square
deviations $D$ as a function of the SOI strength for $ka=19.6$ and $%
\Phi=0.3\Phi_{0}$.}
\label{scattfig2}
\end{figure}
%%%%%%%%%%%%%%%%%%%%%%%%%%%%%%%%%%%%%%%%%%%%%%%%%%%%%%%%%%%%%%%%%%%%%%%%%%%%%%

As shown in Fig.~\ref{scattfig1}, the most general consequence of these
random scattering events is the overall decrease of the conductance.
However, for strong enough disturbance, more interesting effects can be
seen, namely the splitting of the AB peaks which is more pronounced with
increasing $D$. Note that the scattering has the most dramatic effect for
the AB resonances, i.e. $\Phi =n\Phi _{0}$ and the least for the
anti-resonance condition, i.e. $\Phi =\left( n+1/2\right) \Phi _{0}$. Fig.~%
\ref{scattfig2} shows that the conductance has an oscillatory dependence on
the strength of the SOI. Notice that the introduction of scattering reduces
the periodicity with a factor of two. We want to stress that the model we
considered (random elastic scattering processes in single-electron
approximation) is similar to the case when the Altshuler-Aronov-Spivak (AAS)
effect \cite{AAS81} is expected to survive in a single ring. Our results for
a more complex geometry indicate similar physical consequences of the
scattering events: introduction of new peaks in the AB oscillations and
appearance of $\omega _{\text{SO}}/\Omega $ conductance oscillations for a
given value of $\Phi $. This latter effect has recently been predicted for a
single ring \cite{VKN06} and was\ detected in the case of ring arrays.~\cite%
{BKSN06}

\section{Summary}

In this paper we calculated the spin dependent transport properties of
two-dimensional ring arrays. We applied general boundary conditions for the
case of single quantum rings, which allowed the construction of arrays of
such rings as building blocks. The magnetoconductance of two-dimensional
arrays of 3$\times $3,\ 4$\times $4, and 5$\times $5\ quantum rings
exhibited Aharonov-Bohm and Aharonov-Casher oscillations. We also determined
the spin resolved transmission probabilities of the arrays and found significant
spin rotations depending on the SOI strength.
We introduced point-like random scattering centers between the rings, which,
for strong enough disturbance resulted in the splitting of the AB peaks and
the emergence of an oscillatory behavior of the conductance, characteristic
to the SOI strength.

\section*{Acknowledgments}

This work was supported by the Flemish-Hungarian Bilateral Programme, the
Flemish Science Foundation (FWO-Vl), the Belgian Science Policy and the
Hungarian Scientific Research Fund (OTKA) under Contracts Nos.~T48888,
M36803, M045596. P.F.~was supported by a J.~Bolyai grant of the Hungarian
Academy of Sciences. We thank J. S\'{o}lyom for enlightening discussions.

\section*{Appendix}

Here we present the detailed analytic expressions of the scattering problem
for a general two- and four-terminal ring, in which SOI and a perpendicular
magnetic field is present, the latter of which is considered as a
perturbation. As we have shown in Sec.~\ref{modelsec}, it is sufficient to
consider only one input terminal and determine the connection between the
input and output states, i.e. the reflection and transmission matrices,
since the more general boundary condition of having inputs on all terminals
is just a superposition of such cases with an appropriate rotation of the
matrices (see Eqs.~(\ref{3_Ru_Tu})-(\ref{3_Rv_Tv})). Considering $f_{I}$ as
the only input (i.e. $f_{i\neq I}=0$, in Figs. \ref{fig1a} and \ref{fig1c}),
requiring the continuity of the wave functions and applying Griffith
boundary conditions \cite{G53,X92} at the junctions in both cases, we can
obtain the reflection matrices $\hat{R}^{f_{I}}$\ and $\tilde{R}^{f_{I}}$ of
the two-terminal ring, and of the four-terminal ring, respectively. Both can
be written in a form analogous to that of $R^{f_{I}}$\ of the three terminal
case given by Eq. (\ref{R_3_term}) with%
\begin{eqnarray*}
\hat{\varrho}^{\left( \mu \right) } &=&\frac{4k^{2}\!a^{2}}{\hat{y}^{\left(
\mu \right) }}\left\{ \sin (q^{\left( \mu \right) }\gamma _{1})\sin
(q^{\left( \mu \right) }(2\pi -\gamma _{1}))\right. \\
&&\left. +iq^{\left( \mu \right) }\sin (2q^{\left( \mu \right) }\pi
)\right\} ,
\end{eqnarray*}%
and%
\begin{eqnarray*}
\tilde{\varrho}^{\left( \mu \right) }\! &=&\!\frac{2ka}{\tilde{y}^{\left(
\mu \right) }}\!\left\{ \!k^{3}\!a^{3}\!\left[ \cos (2q^{\left( \mu \right)
}\!\pi )\!+\!\cos (2q^{\left( \mu \right) }\!(\pi \!-\!\gamma
_{3}\!+\!\gamma _{2}\!-\!\gamma _{1}))\right. \right. \\
&&\left. \left. -\!\cos (2q^{\left( \mu \right) }\!(\pi \!-\!\gamma
_{3}\!+\!\gamma _{2}))\!+\!\cos (2q^{\left( \mu \right) }\!(\pi \!-\!\gamma
_{3}\!+\!\gamma _{1}))\right. \right. \\
&&\left. \left. -\!\cos (2q^{\left( \mu \right) }\!(\pi \!-\!\gamma
_{2}\!+\!\gamma _{1}))\!-\!\cos (2q^{\left( \mu \right) }\!(\pi \!-\!\gamma
_{3}))\right. \right. \\
&&\left. \left. +\!\cos (2q^{\left( \mu \right) }\!(\pi \!-\!\gamma
_{2}))\!-\!\cos (2q^{\left( \mu \right) }\!(\pi \!-\!\gamma _{1}))\right]
\right. \\
&&\left. +\!2ik^{2}\!a^{2}\!q^{\left( \mu \right) }\!\left[ \sin (2q^{\left(
\mu \right) }\!(\pi \!-\!\gamma _{3}\!+\!\gamma _{2}))\!-\!3\sin (2q^{\left(
\mu \right) }\!\pi )\right. \right. \\
&&\left. \left. +\!\sin (2q^{\left( \mu \right) }\!(\pi \!-\!\gamma
_{3}\!+\!\gamma _{1}))\!+\!\sin (2q^{\left( \mu \right) }\!(\pi \!-\!\gamma
_{2}\!+\!\gamma _{1}))\right] \right. \\
&&\left. +\!4ik^{2}\!a^{2}\!q^{\left( \mu \right) }\!\left[ \sin (2q^{\left(
\mu \right) }\!(\pi \!-\!\gamma _{1}))\!-\!\sin (2q^{\left( \mu \right)
}\!(\pi \!-\!\gamma _{3}))\right] \right. \\
&&\left. -\!4ka(q^{\left( \mu \right) })^{2}\!\left[ \cos (2q^{\left( \mu
\right) }\!(\pi \!-\!\gamma _{3}))\!+\!\cos (2q^{\left( \mu \right) }\!(\pi
\!-\!\gamma _{2}))\right. \right. \\
&&\left. \left. +\!\cos (2q^{\left( \mu \right) }\!(\pi \!-\!\gamma
_{1}))\!-\!3\cos (2q^{\left( \mu \right) }\!\pi )\right] \!\right. \\
&&\left. -\!8i(q^{\left( \mu \right) })^{3}\!\sin (2q^{\left( \mu \right)
}\!\pi )\right\} ,
\end{eqnarray*}%
respectively. Here%
\begin{eqnarray*}
\hat{y}^{\left( \mu \right) }\! &=&\!k^{2}\!a^{2}\left[ \cos (2q^{\left( \mu
\right) }(\pi \!-\!\gamma _{1}))\!-\!\cos (2q^{\left( \mu \right) }\pi )%
\right] \\
&&+4ikaq^{\left( \mu \right) }\sin (2q^{\left( \mu \right) }\pi ) \\
&&-4(q^{\left( \mu \right) })^{2}\!\left[ \cos \!\left[ (\left( -1\right)
^{\mu +1}w\!+\!2\phi )\pi \right] \!+\!\cos (2q^{\left( \mu \right) }\!\pi )%
\right] ,
\end{eqnarray*}%
\begin{eqnarray*}
\tilde{y}^{\left( \mu \right) }\! &=&\!16(q^{\left( \mu \right) })^{4}\!%
\left[ \cos \left[ (\left( -1\right) ^{\mu +1}w\!+\!2\phi )\pi \right]
\!+\!\cos (2q^{\left( \mu \right) }\!\pi )\right] \\
&&-\!32ika(q^{\left( \mu \right) })^{3}\!\sin (2q^{\left( \mu \right) }\!\pi
)\!+\!24k^{2}\!a^{2}(q^{\left( \mu \right) })^{2}\!\cos (2q^{\left( \mu
\right) }\!\pi ) \\
&&-\!4k^{2}\!a^{2}(q^{\left( \mu \right) })^{2}\!\left[ \cos (2q^{\left( \mu
\right) }\!(\pi \!-\!\gamma _{3}))\!+\!\cos (2q^{\left( \mu \right) }\!(\pi
\!-\!\gamma _{2}))\right. \\
&&\left. +\!\cos (2q^{\left( \mu \right) }\!(\pi \!-\!\gamma _{1}))\!+\!\cos
(2q^{\left( \mu \right) }\!(\pi \!-\!\gamma _{3}\!+\!\gamma _{1}))\right. \\
&&\left. +\!\cos (2q^{\left( \mu \right) }\!(\pi \!-\!\gamma _{3}\!+\!\gamma
_{2}))\!+\!\cos (2q^{\left( \mu \right) }\!(\pi \!-\!\gamma _{2}\!+\!\gamma
_{1}))\right] \\
&&-\!8ik^{3}\!a^{3}q^{\left( \mu \right) }\!\sin (2q^{\left( \mu \right)
}\!\pi ) \\
&&+\!4ik^{3}\!a^{3}\!q^{\left( \mu \right) }\!\left[ \!\sin (2q^{\left( \mu
\right) }\!(\pi \!-\!\gamma _{3}\!+\!\gamma _{2}))\!-\!\sin (2q^{\left( \mu
\right) }\!(\pi \!-\!\gamma _{3}))\right. \\
&&\left. +\!\sin (2q^{\left( \mu \right) }\!(\pi -\!\gamma _{2}\!+\!\gamma
_{1}))\!+\!\sin (2q^{\left( \mu \right) }\!(\pi \!-\!\gamma _{1}))\right] \\
&&+\!k^{4}\!a^{4}\left[ \cos (2q^{\left( \mu \right) }\!(\pi \!-\!\gamma
_{3}\!+\!\gamma _{2}\!-\!\gamma _{1}))\!+\!\cos (2q^{\left( \mu \right)
}\!\pi )\right. \\
&&\left. +\!\cos (2q^{\left( \mu \right) }\!(\pi -\!\gamma _{3}\!+\!\gamma
_{1}))\!-\!\cos (2q^{\left( \mu \right) }\!(\pi \!-\!\gamma _{3}\!+\!\gamma
_{2}))\right. \\
&&\left. -\!\cos (2q^{\left( \mu \right) }\!(\pi \!-\!\gamma _{2}\!+\!\gamma
_{1}))\!-\!\cos (2q^{\left( \mu \right) }\!(\pi \!-\!\gamma _{3}))\right. \\
&&\left. +\!\cos (2q^{\left( \mu \right) }\!(\pi \!-\!\gamma _{2}))\!-\!\cos
(2q^{\left( \mu \right) }\!(\pi \!-\!\gamma _{1}))\right] ,
\end{eqnarray*}%
where the angles $\gamma _{i}$ are defined in Fig.~\ref{fig1a} and %
\subref{fig1c}. The transmission matrices $\hat{T}^{f_{I}}$\ of the
two-terminal ring and $\tilde{T}_{n}^{f_{I}}$ ($n=1,2,3$) of the
four-terminal ring, can be given in an analogous form to that of the
transmission matrices $T_{n}^{f_{I}}$ of the three-terminal one given by Eq.
(\ref{T_3_term}) with%
\begin{eqnarray*}
\hat{\tau}^{\left( \mu \right) }\! &=&\!\frac{4ikaq^{\left( \mu \right) }}{%
\hat{y}^{\left( \mu \right) }}e^{i\gamma _{1}\left( \left( -1\right) ^{\mu
+1}w/2+\phi \right) }\left[ \sin (q^{\left( \mu \right) }(2\pi -\gamma
_{1}))\right. \\
&&\left. -e^{-i\pi \left( \left( -1\right) ^{\mu +1}w+2\phi \right) }\sin
(q^{\left( \mu \right) }\gamma _{1})\right] ,
\end{eqnarray*}%
and%
\begin{eqnarray*}
\tilde{\tau}_{1}^{\left( \mu \right) }\! &=&\!\frac{4kaq^{\left( \mu \right)
}}{\tilde{y}^{\left( \mu \right) }}e^{i\gamma _{1}/2\left( \left( -1\right)
^{\mu +1}w+2\phi \right) }\times \\
&&\left\{ ik^{2}\!a^{2}\left[ \sin (q^{\left( \mu \right) }\!(2\pi
\!-\!2\gamma _{3}\!+\!2\gamma _{2}\!-\!\gamma _{1}))\right. \right. \\
&&\left. \left. -\!\sin (q^{\left( \mu \right) }\!(2\pi \!-\!\gamma
_{1}))\!+\!\sin (q^{\left( \mu \right) }\!(2\pi \!-\!2\gamma _{2}\!+\!\gamma
_{1}))\right] \right. \\
&&\left. \left. -\!\sin (q^{\left( \mu \right) }\!(2\pi \!-\!2\gamma
_{3}\!+\!\gamma _{1}))\right] \right. \\
&&\left. -\!2kaq^{\left( \mu \right) }\!\left[ \cos (q^{\left( \mu \right)
}\!(2\pi \!-\!2\gamma _{2}\!+\!\gamma _{1}))\right. \right. \\
&&\left. \left. -\!2\cos (q^{\left( \mu \right) }\!(2\pi \!-\!\gamma
_{1}))\!+\!\cos (q^{\left( \mu \right) }\!(2\pi \!-\!2\gamma _{3}\!+\!\gamma
_{1}))\right] \right. \\
&&\left. +\!4i(q^{\left( \mu \right) })^{2}\!\left[ e^{-i\pi \left( \left(
-1\right) ^{\mu +1}w+2\phi \right) }\sin (q^{\left( \mu \right) }\!\gamma
_{1})\right. \right. \\
&&\left. \left. -\!\sin (q^{\left( \mu \right) }\!(2\pi \!-\!\gamma _{1})) 
\right] \right\} ,
\end{eqnarray*}%
\begin{eqnarray*}
\tilde{\tau}_{2}^{\left( \mu \right) }\! &=&\!\frac{4kaq^{\left( \mu \right)
}}{\tilde{y}^{\left( \mu \right) }}e^{i\gamma _{2}/2\left( \left( -1\right)
^{\mu +1}w+2\phi \right) }\times \\
&&\left\{ -2kaq^{\left( \mu \right) }\!\left[ e^{-i\pi \left( \left(
-1\right) ^{\mu +1}w+2\phi \right) }\cos (q^{\left( \mu \right) }\!\gamma
_{2})\right. \right. \\
&&\left. \left. -\!e^{-i\pi \left( \left( -1\right) ^{\mu +1}w+2\phi \right)
}\cos (q^{\left( \mu \right) }\!(2\gamma _{1}\!-\!\gamma _{2}))\right.
\right. \\
&&\left. \left. +\!\cos (q^{\left( \mu \right) }\!(2\pi \!-\!2\gamma
_{3}\!+\!\gamma _{2}))\!-\!\cos (q^{\left( \mu \right) }\!(2\pi \!-\!\gamma
_{2}))\right] \right. \\
&&\left. +\!4i(q^{\left( \mu \right) })^{2}\!\left[ e^{-i\pi \left( \left(
-1\right) ^{\mu +1}w+2\phi \right) }\sin (q^{\left( \mu \right) }\!\gamma
_{2})\right. \right. \\
&&\left. \left. -\!\sin (q^{\left( \mu \right) }\!(2\pi \!-\!\gamma _{2})) 
\right] \right\} ,
\end{eqnarray*}%
\begin{eqnarray*}
\tilde{\tau}_{3}^{\left( \mu \right) } &=&\frac{4kaq^{\left( \mu \right) }}{%
\tilde{y}^{\left( \mu \right) }}e^{i\gamma _{3}/2\left( \left( -1\right)
^{\mu +1}w+2\phi \right) }\times \\
&&\left\{ ik^{2}\!a^{2}e^{-i\pi \left( \left( -1\right) ^{\mu +1}w+2\phi
\right) }\left[ \sin (q^{\left( \mu \right) }\!\gamma _{3})\right. \right. \\
&&\left. \left. +\sin (q^{\left( \mu \right) }\!(2\gamma _{1}\!-\!\gamma
_{3}))\!-\!\sin (q^{\left( \mu \right) }\!(2\gamma _{2}\!-\!\gamma
_{3}))\right. \right. \\
&&\left. \left. +\sin (q^{\left( \mu \right) }\!(2\gamma _{2}\!-\!2\gamma
_{1}\!-\!\gamma _{3}))\right] \right. \\
&&\left. -2kaq^{\left( \mu \right) }e^{-i\pi \left( \left( -1\right) ^{\mu
+1}w+2\phi \right) }\left[ 2\cos (q^{\left( \mu \right) }\!\gamma
_{3})\right. \right. \\
&&\left. \left. -\cos (q^{\left( \mu \right) }\!(2\gamma _{1}\!-\!\gamma
_{3}))\!-\!\cos (q^{\left( \mu \right) }\!(2\gamma _{2}\!-\!\gamma _{3})) 
\right] \right. \\
&&\left. +4i(q^{\left( \mu \right) })^{2}\!\left[ e^{-i\pi \left( \left(
-1\right) ^{\mu +1}w+2\phi \right) }\sin (q^{\left( \mu \right) }\!\gamma
_{3})\right. \right. \\
&&\left. \left. -\sin (q^{\left( \mu \right) }\!(2\pi \!-\!\gamma _{3})) 
\right] \right\} ,
\end{eqnarray*}%
respectively.

\end{document}